\documentclass[superscriptaddress,aps,pra,nofootinbib,notitlepage,10pt,longbibliography, reprint, onecolumn]{revtex4-1}
\pdfoutput=1
\usepackage{graphicx}
\usepackage{amsmath}
\usepackage{amssymb}
\usepackage{amsthm}
\usepackage{comment}
\usepackage{cprotect}
\usepackage{placeins}
\usepackage[caption=false]{subfig}
\usepackage[colorlinks]{hyperref}
\usepackage[all]{hypcap}
\usepackage{tikz}
\usepackage{verbatim}
\usepackage{subfig}
\usetikzlibrary{arrows}
\usepackage{units}
\usepackage{soul}
\usepackage{textcomp}
\usepackage{float}
\usepackage{algorithmic}
\usepackage[version=4]{mhchem}

\usepackage{listings}
\usepackage{color}
\usepackage[utf8]{inputenc}
\definecolor{codegreen}{rgb}{0,0.6,0}
\definecolor{codegray}{rgb}{0.5,0.5,0.5}
\definecolor{codepurple}{rgb}{0.58,0,0.82}
\definecolor{backcolour}{rgb}{0.95,0.95,0.92}

\lstdefinestyle{mystyle}{
  backgroundcolor=\color{backcolour},   commentstyle=\color{codegreen},
  keywordstyle=\color{magenta},
  numberstyle=\tiny\color{codegray},
  stringstyle=\color{codepurple},
  basicstyle=\footnotesize,
  breakatwhitespace=false,
  breaklines=true,
  captionpos=b,
  keepspaces=true,
  numbers=left,
  numbersep=5pt,
  showspaces=false,
  showstringspaces=false,
  showtabs=false,
  tabsize=2
}

\newcommand{\eq}[1]{Eq.~\hyperref[eq:#1]{(\ref*{eq:#1})}}
\renewcommand{\sec}[1]{\hyperref[sec:#1]{Section~\ref*{sec:#1}}}
\DeclareRobustCommand{\app}[1]{\hyperref[app:#1]{Appendix~\ref*{app:#1}}}
\newcommand{\tab}[1]{\hyperref[tab:#1]{Table~\ref*{tab:#1}}}
\newcommand{\fig}[1]{\hyperref[fig:#1]{Figure~\ref*{fig:#1}}}
\newcommand{\figa}[2]{\hyperref[fig:#1]{Figure~\ref*{fig:#1}#2}}
\newcommand{\figx}[2]{\hyperref[fig:#1]{Figure~\ref*{fig:#1}(#2)}}
\newcommand{\thm}[1]{\hyperref[thm:#1]{Theorem~\ref*{thm:#1}}}
\newcommand{\lem}[1]{\hyperref[lem:#1]{Lemma~\ref*{lem:#1}}}
\newcommand{\cor}[1]{\hyperref[cor:#1]{Corollary~\ref*{cor:#1}}}
\newcommand{\defn}[1]{\hyperref[def:#1]{Definition~\ref*{def:#1}}}
\newcommand{\alg}[1]{\hyperref[alg:#1]{Algorithm~\ref*{alg:#1}}}

\input{Qcircuit}

\begin{document}

\title{Compressing Many-Body Fermion Operators Under Unitary Constraints}

\date{\today}

\author{Nicholas C.~Rubin}
\email{nickrubin@google.com}
\affiliation{Google Quantum AI, Mountain View CA}
\author{Joonho Lee}
\email{linusjoonho@gmail.com} 
\affiliation{Google Quantum AI, Mountain View CA}
\affiliation{Department of Chemistry, Columbia University, New York, New York 10027, USA}
\author{Ryan Babbush}
\email{ryanbabbush@gmail.com} 
\affiliation{Google Quantum AI, Mountain View CA}

\begin{abstract}
The most efficient known quantum circuits for preparing unitary coupled cluster states and applying Trotter steps of the arbitrary basis electronic structure Hamiltonian involve interleaved sequences of fermionic Gaussian circuits and Ising interaction type circuits. These circuits arise from factorizing the two-body operators generating those unitaries as a sum of squared one-body operators that are simulated using product formulas. We introduce a numerical algorithm for performing this factorization that has an iteration complexity no worse than single particle basis transformations of the two-body operators and often results in many times fewer squared one-body operators in the sum of squares compared to the analytical decompositions. As an application of this numerical procedure, we demonstrate that our protocol can be used to approximate generic unitary coupled cluster operators and prepare the necessary high-quality initial states for techniques (like ADAPT-VQE) that iteratively construct approximations to the ground state.
\end{abstract}

\maketitle
\section{Introduction}
Efficient quantum circuit compilation is an important task for performing quantum simulations on near-term and fault-tolerant quantum devices. Different approximation schemes can lead to vastly different circuit representations and thus varying runtimes and success probabilities~\cite{hastings2014improving, babbush2015chemical, kivlichan2018quantum, tubman2018postponing}.  In this work we focus on the general problem of circuit implementations of propagators generated by fermionic many-body operators relevant to a wide variety of simulation such as time-dynamics and state preparation in electronic structure simulations of chemistry and condensed matter models. 

The strategy we explore is the decomposition of a generic two-body operator into a sum-of-squares of normal operators where each term in the sum can be implemented exactly with fermionic Gaussian unitaries (i.e., single-particle rotations) and unitaries generated by charge-charge (i.e., Ising) type interactions.  Circuits of this form can be implemented exactly through Givens rotation networks~\cite{PhysRevLett.120.110501} and swap networks~\cite{o2019generalized, kivlichan2018quantum}.  Many recently proposed simulation strategies for fermions leverage an analytical sum-of-squares many-body operator decomposition or use a sum-of-squares type ansatz for approximate ground states.  Some examples for both near-term quantum computers and fault tolerant quantum computers are the double factorized Trotter steps for chemical Hamiltonians~\cite{motta2018low}, tensor-hypercontraction based hamiltonian dynamics~\cite{lee2020even}, restricted models of generalized coupled-cluster~\cite{lee2018generalized,  matsuzawa2020jastrow, kottmann2021optimized}, and compressed density fitting~\cite{cohn2021quantum}.
The sum-of-squares picture unifies these ans\"atze and suggests a numerical compilation strategy for determining a sum-of-squares operator decomposition with few terms. In the context of coherence limited near-term quantum computers we highlight how a numerical approach can lead to substantially shorter circuits for the implementation of many-body operators relevant to quantum chemistry.

A numerical optimization strategy for a non-orthogonal single particle basis representation of many-body operators has already been used in determining one-particle bases to measure chemical Hamiltonians~\cite{yen2020cartan} and compressing many-body operators through numerical density fitting~\cite{cohn2021quantum}.  A classical analog of these methods is matching pursuit where the dictionary is a set of non-orthogonal single particle bases which are obtained as the algorithm progresses~\cite{wu2003matching, mcclean2015compact}.  In the context of many-body operators, these methods are connected by the Lie algebraic perspective on operator decomposition into bases that maximize their Cartan subalgebra representations~\cite{yen2020cartan}.  Numerically determining a unitary that maximizes the Cartan subalgebra of a qubit operator has also been used to determine efficient compilations for two-local qubit Hamiltonians~\cite{kokcu2021fixed}.  A missing component of many of these proposals is an efficient computational scheme for determining the basis that maximizes the Cartan subalgebra components.  

We propose an efficient local search (i.e., greedy) algorithm for recursively decomposing a fermionic two-body operator into a sum-of-squares terms with an iteration cost that scales no worse than a one-particle basis rotation of the two-body operator--$\mathcal{O}(n^{5})$ where $n$ is the number of spin-orbitals describing the problem.  A numerical study on decomposition of the unitary coupled cluster operator indicates that a greedy search provides substantial improvements over analytical decompositions such as the singular value decomposition~\cite{motta2018low} or Takagi~\cite{matsuzawa2020jastrow} decomposition and provides similar performance to least-squares tensor fitting~\cite{cohn2021quantum} at substantially reduced computational complexity.   

To further highlight the utility of this approach we demonstrate how compressed unitary coupled cluster doubles can serve as a starting point for iterative wavefunction methods that rely on the initial wavefunction overlap with the ground state to succeed. In classical quantum chemistry, iterative approaches based on similarity transformations of a state~\cite{yanai2006canonical, white2002numerical} or reduced-density-matrix propagation~\cite{PhysRevLett.97.143002} require high accuracy initial states if a high quality approximations to the true ground state is desired.  We validate that quantum versions of these techniques, such as ADAPT-VQE~\cite{grimsley2018adapt} or the quantum antihermitian-contracted Schr\"odinger equation solver~\cite{PhysRevLett.126.070504}, can also be sensitive to initial states by studying the performance of ADAPT-VQE on finding a high accuracy approximation to the ground state of \ce{O2}. This example highlights how iterative circuit construction techniques can fail in the absence of high initial overlap with the target state, even when one introduces artificial symmetry breaking. 

In Section~\ref{sec:sos_derivation} we describe the sum-of-squares decomposition of a generic fermionic two-body operator and how this translates into the Gaussian unitary and Ising swap network circuit primitives under a Trotter approximation. In Section~\ref{sec:greedy_opt} we describe our greedy algorithm for performing the decomposition of a generic two-body operator into a sum-of-squares form.  In Section~\ref{sec:results} we compare the numerical sum-of-squares decomposition to analytical techniques for unitary coupled-cluster generators along with an application to iterative circuit constructions.  We close with perspectives on the numerical compression and when it is most applicable in the context of quantum computing for simulating fermions.  

\section{A Sum-of-squared normal operator representation}\label{sec:sos_derivation}

\subsection{Background on simulating two-body fermion operators as a sum of squared one-body operators}
Starting from a generic antihermitian two-body operator
\begin{align}\label{eq:generic_g}
G =& \sum_{pqrs}A_{rs}^{pq}a_{p}^{\dagger}a_{q}^{\dagger}a_{s}a_{r}
\end{align}
where $\{p,q,r,s\}$ index fermionic modes, the charge-charge form can be obtained by reordering ladder operators under the fermionic anticommutation relations
\begin{align}
G =& \sum_{pq,rs}A^{pq}_{sr}\left(a_{p}^{\dagger}a_{s}a_{q}^{\dagger}a_{r} - \delta_{s}^{q}a_{p}^{\dagger}a_{r}\right)
\end{align}
where the $A \in \mathbb{C}^{\times^{n}}$ and is antisymmetric in the upper and lower indices. A sum-of-squares of normal operator decomposition of $G$ has the form 
\begin{align}\label{eq:sos_op}
 G = \sum_{l}Z_{l}^{2} -  \sum_{pr}S_{pr}a_{p}^{\dagger}a_{r} \;\;,\;\; Z_{l}= \sum_{pq}z(l)_{pq}a_{p}^{\dagger}a_{q}
\end{align}
where $z(l)_{pq}$ is a collection of coefficients such that $Z_{l}$ is a normal operator and $S_{pr} = \sum_{qs}A_{rs}^{pq}\delta_{s}^{q}$. 
Ans\"atze generated by $G$ can be viewed as a unitary form of the generalized coupled-cluster ansatz of Nooijen~\cite{PhysRevLett.84.2108} or, if $A$ is a imaginary antihermitian tensor with the correct symmetries, a generic quantum chemistry Hamiltonian evolution.  

Under the exponential map the operator $G$ expressed in the form of Eq~\eqref{eq:sos_op} admits a simple compilation strategy through a first order Trotter approximation
\begin{align}\label{eq:approx_lr_evolution_unitary}
e^{G} \approx e^{-S}\prod_{l}e^{Z_{l}^{2}}.
\end{align}
By representing each $Z_{l}$ in its eigenbasis each $Z_{l}^{2}$ term in Eq.~\eqref{eq:approx_lr_evolution_unitary} can be exactly implemented as 
\begin{align}\label{eq:unitary_squared_normal}
e^{Z_{l}^{2}} = U_{l} e^{\sum_{pq}J_{pq}(l)n_{p}n_{q}} U_{l}^{\dagger}
\end{align}
where $U_{l}$ is a fermionic Gaussian rotating to the eigenbasis of $Z_{l}$ and $J(l)_{pq}$ corresponds to the outer product of eigenvalues  of $z(l)$.
Equation~\ref{eq:approx_lr_evolution_unitary} becomes
\begin{align}
e^{G} \approx e^{-\sum_{pq}S_{pq}a_{p}^{\dagger}a_{q}}U^{\dagger}_{L}\prod_{l=1}^{L}e^{\sum_{pq}J_{pq}(l)n_{p}n_{q}}\tilde{U}_{l}
\end{align}
where $\tilde{U}_{l} = U_{l}U_{l-1}^{\dagger}$ is the concatenation of single-particle basis rotations--we take $U_{0}$ to be identity.  This concatenation can be performed classically and implemented on the quantum computer as a linear depth Givens rotation network~\cite{PhysRevLett.73.58, clements2016optimal, PhysRevApplied.9.044036, PhysRevLett.120.110501, rubinScience2020}.  The charge-charge component of this sequence of unitaries $e^{\sum_{pq}J_{pq}(l)n_{p}n_{q}}$ can be implemented with a linear depth swap network~\cite{o2019generalized, kivlichan2018quantum}.

The Trotterized form of $e^{G}$ is thus implemented as a sequence of linear depth circuits alternating between Givens rotations networks and charge-charge type networks all of which require only nearest-neighbor connectivity between a linear array of qubits. Furthermore, these compilations are conjectured to be optimal for these decompositions.  The scaling of this implementation is linear in the rank of the matrix $\tilde{A}$ which is the super matrix formed from reshaping $A$ such that the row and column indices are labeled by pair indices $ps$ and $qr$ (see Appendix~\ref{app:alternative_decomps} for a detailed discussion on geminal ordering of the supermatrix).  
\subsection{Determining normal operators}
There are a variety of methods for decomposing $G$ into sum-of-squared normal operators as Eq~\eqref{eq:sos_op}.  In Reference~\cite{motta2018low} the double factorization technique, similar to the Cholesky decomposition, is applied to the $\tilde{A}$ super matrix to produce a sum-of-squares representation.  The factors from the Cholesky decomposition can be reshaped and factorized again, via an eigenvalue decomposition, because of the four fold symmetry of the Hamiltonian coefficients coming from the two-electron integral coefficients. Reference~\cite{motta2018low} also demonstrated a general decomposition of a unitary coupled cluster operator that relies on a singular value decomposition (SVD) of $\tilde{A}$. In Appendix~\ref{app:alternative_decomps} we derive the SVD decomposition for an arbitrary operator $G$ without the structural requirements of a coupled cluster doubles operator.  A slightly more efficient decomposition for unitary coupled-cluster doubles operators is pointed out by Mastsuzawa \textit{et al.}~\cite{matsuzawa2020jastrow} in the context of implementing Jastrow inspired sum-of-squares many-body operators by leveraging the Takagi decomposition~\cite{PhysRevA.94.062109}.  The Takagi decomposition is applicable to complex symmetric matrices as the decomposition
\begin{align}\label{eq:takagi_decomp}
\tilde{A} = U \mathrm{diag}(\sigma_{1},...,\sigma_{m}) U^{T} 
\end{align}
where $\sigma_{i} \geq 0$ and $U$ is unitary. In general the Takagi decomposition is not equivalent to a scaled SVD~\cite{hahn2006routines} as $U$ is complex.  To form a sum-of-squares decomposition from the Takagi decomposition we follow Reference~\cite{matsuzawa2020jastrow} and reshape the columns of $U$ into matrices $u(l)$ where $l$ indexes the column and define
\begin{align}
y(l) = \sqrt{\sigma(l)}u(l)
\end{align}
such that
\begin{align}
A^{ps}_{qr} = \sum_{l}y(l)_{ps}y(l)_{qr}.
\end{align}
Because $y(l)$ is not generally a normal matrix we can form one by taking linear combinations 
\begin{align}
y(l)^{\pm} = y(l) \pm i y(l)^{\dagger}.
\end{align}
 such that
\begin{align}
A^{ps}_{qr} = \frac{1}{4}\sum_{l}\left( y(l)^{+}_{ps}y(l)^{+}_{qr} + y(l)^{-}_{ps}y(l)^{-}_{qr}\right).
\end{align}
The coefficients of $y(l)^{\pm}$ can be thought of as the coefficients for the one-body operators $z(l)$ in Eq.~\eqref{eq:sos_op}. Thus they can be diagonalized by a unitaries $\mu^{\pm}$ such that 
\begin{align}\label{eq:charge_charge_a}
A^{ab}_{cd} = \frac{1}{4}\sum_{l,pq}\left(\mu^{+*}(l)_{ap}\mu^{+}_{cp}(l)J_{pq}^{+}(l)\mu^{+*}_{bq}(l)\mu^{+}_{dq}(l) + \mu^{-*}(l)_{ap}\mu^{-}_{cp}(l)J_{pq}^{-}(l)\mu^{-*}_{bq}(l)\mu^{-}_{dq}(l)  \right)
\end{align}
where $J_{pq}^{\pm}(l) = \lambda_{p}^{\pm}(l)\lambda_{q}^{\pm}(l)$ and $\lambda_{p}^{\pm}(l)$ come from diagonalizing $y(l)^{\pm}$. A full derivation of this general form is described in Appendix~\ref{app:alternative_decomps}. Written more abstractly
\begin{align}\label{eq:abstract_sos}
A^{ab}_{cd} = \sum_{l}\mu(l)^{*}_{ap}\mu(l)_{cp}J_{pq}(l)\mu(l)_{bq}^{*}\mu(l)_{dq}
\end{align}
where we have moved the $\pm$ terms into the $l$ sum.  Both the Takagi decomposition and the SVD sum-of-squares decomposition form an operator decomposition like Eq.~\eqref{eq:abstract_sos}.
It was previously pointed out that $J_{pq}(l)$ is a rank-one matrix, and thus carries little information~\cite{matsuzawa2020jastrow}. This justifies a numerical optimization of this doubles term, as is performed in the unitary cluster Jastrow factor ansatz~\cite{matsuzawa2020jastrow} ($k$-uCJ) and its variants~\cite{lee2018generalized, kottmann2021optimized}.  Reference~\cite{cohn2021quantum} proposed an approach based on a gradient descent least-squares fitting of $\mu(l)$ and $J(l)$ for $\tilde{A}$ under the constraint that the coefficients have the proper symmetry to represent a spin-free chemical Hamiltonian. Although Reference~\cite{cohn2021quantum} introduces a clever optimization strategy that alternates between one particle basis $\mu(l)$ optimization and $J(l)$ coefficient optimization, the overall convergence of the the least-squares fitting is unknown and seems to be limited to six tensor factors before numerical difficulties make optimization challenging. In Section~\ref{sec:greedy_opt} we propose a fitting algorithm that determines a decomposition of $A$ according to Eq~\eqref{eq:abstract_sos} which has comparable performance but is substantially more computationally efficient.

\subsection{$S_{z}$-Symmetry Adaptation}
\begin{figure}
    \centering
    \includegraphics[width=8.5cm]{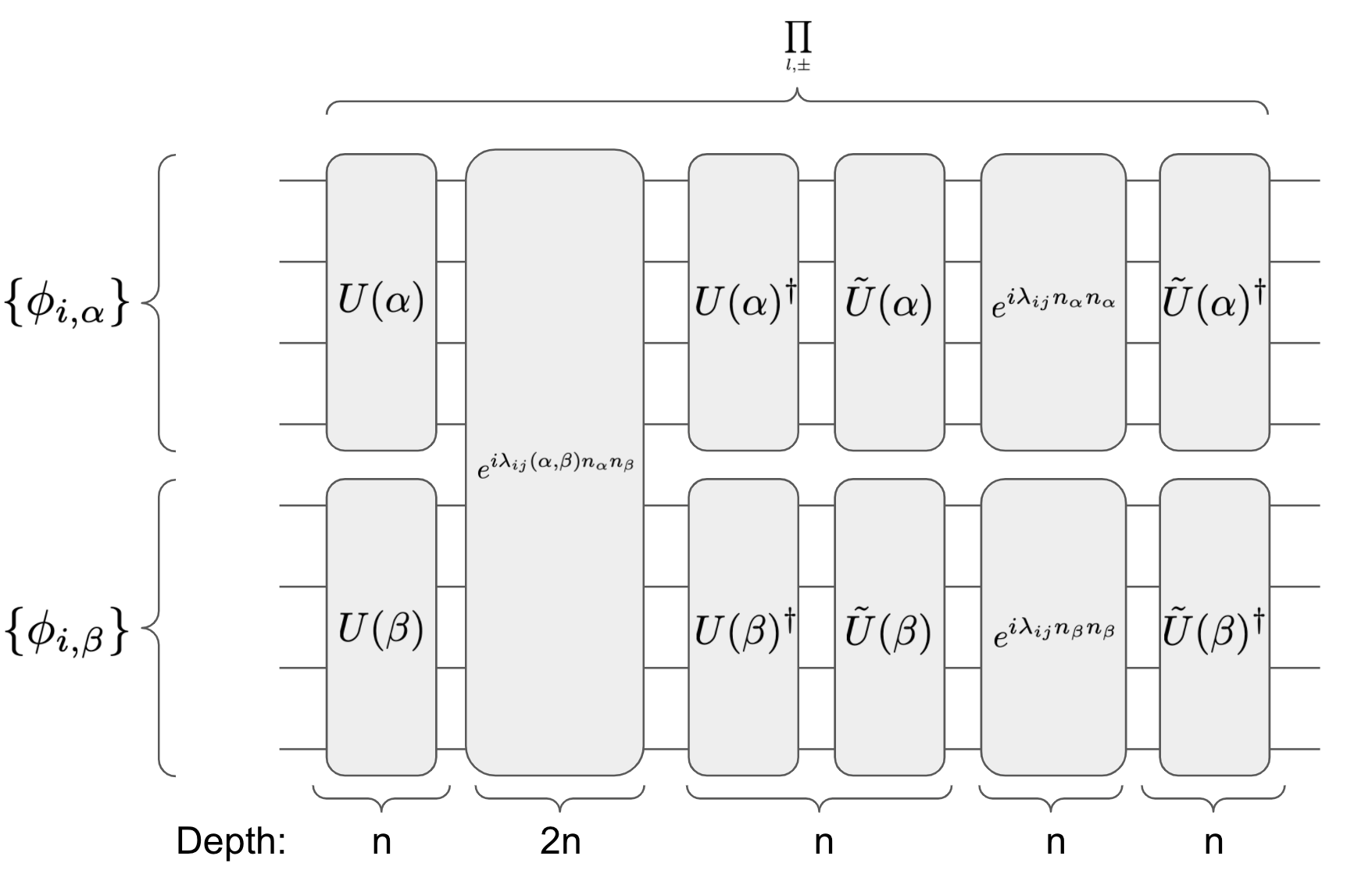}
    \caption{Depiction of a single Trotter slice of the sum-of-squares decomposition for each spin sector. Each $U(\tau)$ is implemented as 
    a linear depth Givens rotation circuit and each $e^{i\sum_{ij}\lambda_{ij}n_{i}n_{j}}$ is also 
    implemented in a linear depth swap network pattern. Each swap network or basis change unitary consists of $\mathcal{O}(n^{2})$ gates.}
    \label{fig:takagi_circuit_spin_sector}
\end{figure}

Both the SVD scheme and Takagi scheme can be implemented in such a fashion that all generators commute with $\hat{S}_{z}$\footnote{For fermionic systems $\hat{S}_{z} = \frac{1}{2}\sum_{n}\left(a_{n\alpha}^{\dagger}a_{n\alpha} - a_{n\beta}^{\dagger}a_{n\beta}\right)$ where the sum is over the spatial component of the fermionic modes and the $\{\alpha, \beta\}$ index indicates the spin of the spin-orbital.}. Specifically, this can be achieved by applying the decompositions to $\tilde{A}$ partitioned into the non-redundant spin components $\tilde{A}^{\alpha,\alpha}_{\alpha,\alpha}$,  $\tilde{A}^{\alpha,\beta}_{\alpha,\beta}$,  and $\tilde{A}^{\beta,\beta}_{\beta,\beta}$.  Consider the spin-indexed generator $\tilde{G}$,
\begin{align}
\tilde{G}_{p\sigma s\sigma, q\tau r\tau} = a_{p\sigma}^{\dagger}a_{s\sigma}a_{q\tau}^{\dagger}a_{r\tau} - a_{p\sigma}^{\dagger}a_{r\tau}\delta_{q\tau}^{s\sigma}.
\end{align}
We can view the spin-block structure in $\tilde{A}$ in matrix form (via the same supermatrix formed in the Takagi and SVD decomposition) 
\begin{align}
\tilde{A} = 
\begin{pmatrix}
\tilde{A}(\alpha, \alpha) & \tilde{A}(\alpha, \beta) \\
\tilde{A}(\beta, \alpha) & \tilde{A}(\beta, \beta)
\end{pmatrix}
 = 
 \begin{pmatrix}
 A & B \\
 B^{T} & C
 \end{pmatrix}
\end{align}
which is a complex symmetric matrix. $\tilde{A}(\tau, \sigma)$ indicates all terms with two-body operators of the form $a_{p\tau}^{\dagger}a_{q\tau}a_{r\sigma}^{\dagger}a_{s\sigma}$. By performing one of the sum-of-squares decomposition on the $A$ and $C$ blocks and then the larger matrix involving $B$ and $B^{T}$ we are guaranteed the one-particle basis rotations for each sum-of-squares operator is restricted to a single spin-sector. The $A$ and $C$ blocks can be implemented simultaneously and can be merged with the single-particle basis transformation obtained by rearranging $a_{p}^{\dagger}a_{q}^{\dagger}a_{r}a_{s} \rightarrow a_{p}^{\dagger}a_{s}a_{q}^{\dagger}a_{r}$. Figure~\ref{fig:takagi_circuit_spin_sector} shows an example of the circuit compilation for one of the $l$ terms in the Takagi or SVD decomposition of $\tilde{A}$ that commutes with $S_{z}$. 
\section{Compressing Many-Body Operators with Sequential Orbital Optimization}\label{sec:greedy_opt}
The previous section reviewed methods that allow for a coefficient tensor associated with a generic operator to be decomposed into a sum-of-squares of normal operators. It also explained how the evolution by these sum-of-squares operators can be efficiently implemented using fermionic Gaussian unitaries and Ising interactions. 
Here we describe a numerical algorithm to express $A$ in a sum-of-squared normal matrices according to Eq.~\eqref{eq:abstract_sos}. The procedure is a greedy algorithm which sequentially finds a single-particle basis such that $A$ has large coefficients for $n_{i}n_{j}$ terms.  We then remove this component leaving a remainder tensor.  The protocol is repeated until the remainder is numerically zero or the remainder norm is below a preset threshold.  We note that this procedure leads to  matrices $J_{pq}(l)$ (the coefficients associated with the $n_{p}n_{q})$ in Eq~\eqref{eq:unitary_squared_normal}) which are not restricted to being rank one.  We numerically determine the single-particle basis that maximizes $J_{pq}(l)$ through gradient descent on the non-redundant generator coefficients.  Derivatives of the basis rotation unitaries with respect to the generating coefficients are provided in~\cite{wilcox1967exponential, helgaker2014molecular} and have been used in gradient optimization of the Hartree-Fock equations~\cite{rubinScience2020, helgaker2014molecular}.

To derive a cost function for maximizing $J_{pq}(l)$ we consider a one-body transform of a two-body operator
\begin{align}
\hat{T} =& \sum_{ijkl}t_{ij,kl}a_{i}^{\dagger}a_{j}a_{k}^{\dagger}a_{l}
\end{align}
that generates unitary dynamics. Given the single particle basis transformation operator
\begin{align}
U(\kappa) = e^{\sum_{pq}\kappa_{pq}a_{p}^{\dagger}a_{q}}
\end{align}
the orbital rotated generator is represented below as $\tilde{T}$
\begin{align}
\tilde{T} =& \hat{U}^\dagger(\kappa)\hat{T}\hat{U}(\kappa) \\
\tilde{T} &= \sum_{pqrs}\tilde{t}_{pq,rs}a_{p}^{\dagger}a_{q}a_{r}^{\dagger}a_{s} \\
\tilde{t}_{pq,rs} &= \sum_{ijkl}u_{pi}^{*}u_{qj}u_{rk}^{*}u_{sl}t_{ijkl}  \label{eq:tilde_t_transformed}
\end{align}
such that the objective of maximizing the coefficients of the $n_{i}n_{j}$ component of $\tilde{T}$ can be expressed as
\begin{align}\label{eq:opt_max_kappa}
\max_{\kappa} O(\kappa) \; , \; O(\kappa) = \sum_{xy}\vert \tilde{t}_{xx,yy}(\kappa) \vert^{2}.
\end{align}
To optimize, we first take the gradient of the objective function which yields
\begin{align}
\frac{\partial O(\kappa)}{\partial \kappa_{a,b}} = \sum_{xy}2 \mathrm{Re}\left[ \tilde{t}_{xx,yy}\frac{\partial \tilde{t}_{xx,yy}}{\partial \kappa_{a,b}}\right]
\end{align}
To calculate the partial derivative of $\tilde{t}$ coefficients with respect to the parameters of the generator $\kappa$ we first derive the form of $\frac{\partial \tilde{T}}{\partial \kappa_{a,b}}$ in the following equations
\begin{widetext}
\begin{align}
\frac{\partial \tilde{T}}{\partial \kappa_{a,b}} =& \sum_{ijkl}t_{ij,kl}\left(\frac{\partial \hat{U}(\kappa)^{\dagger}}{\kappa_{ab}}a_{i}^{\dagger}a_{j}a_{k}^{\dagger}a_{l}\hat{U}(\kappa) +  \hat{U}(\kappa)^{\dagger}a_{i}^{\dagger}a_{j}a_{k}^{\dagger}a_{l}\frac{\partial \hat{U}(\kappa)}{\partial \kappa_{a,b}} \right) \\
=& \sum_{ijkl}t_{ij,kl}\hat{U}(\kappa)^{\dagger} \left[a_{i}^{\dagger}a_{j}a_{k}^{\dagger}a_{l}, \sum_{mn}W_{mn}^{ab}a_{m}^{\dagger}a_{n} \right] \hat{U}(\kappa) \\
=& \sum_{ijklr}t_{ij,kl}\hat{U}(\kappa)^{\dagger} \left(-W_{ri}^{ab}a_{r}^{\dagger}a_{j}a_{k}^{\dagger}a_{l} + W_{jr}^{ab}a_{i}^{\dagger}a_{r}a_{k}^{\dagger}a_{l} - W_{rk}^{ab}a_{i}^{\dagger}a_{j}a_{r}^{\dagger}a_{l} + W_{lr}^{ab}a_{i}^{\dagger}a_{j}a_{k}^{\dagger}a_{r}\right) \hat{U}(\kappa) \\
=& \sum_{pqst}V_{pq,st}a_{p}^{\dagger}a_{q}a_{s}^{\dagger}a_{t} 
\end{align}
where \begin{align}
V_{pq,st} =& \sum_{ijklr}t_{ij,kl}\left(-W_{ri}^{ab}u_{rp}^{*}u_{jq}u_{ks}^{*}u_{lt} + W_{jr}^{ab}u_{ip}^{*}u_{rq}u_{ks}^{*}u_{lt} - W_{rk}^{ab}u_{ip}^{*}u_{jq}u_{rs}^{*}u_{lt} + W_{lr}^{ab}u_{ip}^{*}u_{jq}u_{ks}^{*}u_{rt} \right)
\end{align}
which uses
\begin{align}
U(\kappa)^{\dagger}a_{i}^{\dagger}U(\kappa) = \sum_{p}(e^{\kappa})_{ip}^{*}a_{p}^{\dagger} = \sum_{p}u_{ip}^{*}a_{p}^{\dagger}\\
U(\kappa)^{\dagger}a_{i}U(\kappa) = \sum_{p}(e^{\kappa})_{ip}a_{p} = \sum_{p}u_{ip}a_{p}
\end{align}
and the matrix $W^{ab}$ are the coefficients for the antihermitian operator obtained from the Wilcox formula~\cite{wilcox1967exponential} which is an analytical expression for the derivative of a unitary with respect to its generating parameter $\kappa_{a,b}$ (see Appendix G of Reference~\cite{rubinScience2020} for a full derivation).  This analytical formula merely requires diagonalizing the generator matrix $\kappa$ and requires no truncation of the matrix exponential Taylor expansion. Therefore,
\begin{align}
\frac{\partial \tilde{t}_{pq,st}}{\partial \kappa_{ab}} =  \sum_{ijklr}t_{ij,kl}\left(-W_{ri}^{ab}u_{rp}^{*}u_{jq}u_{ks}^{*}u_{lt} + W_{jr}^{ab}u_{ip}^{*}u_{rq}u_{ks}^{*}u_{lt} - W_{rk}^{ab}u_{ip}^{*}u_{jq}u_{rs}^{*}u_{lt} + W_{lr}^{ab}u_{ip}^{*}u_{jq}u_{ks}^{*}u_{rt} \right)\label{eq:n7_scaling_deriv}
\end{align}
\end{widetext}
which is a contraction that can be evaluated in $\mathcal{O}(n^{5})$ where $n$ is the number of spin-orbitals.  There are a total of $n^{2}$ such $\kappa_{ab}$ terms when considering the real and imaginary components and thus the total derivative is obtained in $\mathcal{O}(n^{7})$ operations.  Directly evaluating Eq~\eqref{eq:n7_scaling_deriv} is needlessly expensive.  Shown below is a procedure for obtaining the same expression for $\frac{\partial \tilde{t}_{pq,st}}{\partial\kappa_{a,b}}$ in $\mathcal{O}(n^{5})$ operations.

This method of obtaining the gradient is general and sufficient for any continuous cost function depending on $\tilde{t}_{pq,st}$. The scaling to evaluate the gradient can be reduced by considering the function we are optimizing and applying the chain rule.  To see how this works consider differentiating Eq.~\eqref{eq:tilde_t_transformed} with respect to $\kappa_{ab}$.
\begin{align}
\frac{\partial O(\kappa)}{\partial \kappa_{ab}} =& \sum_{cd}\frac{\partial O(\kappa)}{\partial u_{cd}}\left(\frac{\partial u}{\partial \kappa_{ab}}\right)_{cd} + \sum_{cd}\frac{\partial O(\kappa)}{\partial u_{cd}^{*}}\left( \frac{\partial u^{*}}{\partial \kappa_{a,b}}\right)_{cd}
\end{align}
where 
\begin{align}
\frac{\partial O(\kappa)}{\partial u_{cd}} = 2\sum_{xy}\mathrm{Re}\left[\tilde{t}_{xx,yy}\frac{\partial \tilde{t}_{xx,yy}}{\partial u_{cd}}\right] = 2\sum_{yikl}\mathrm{Re}\left[t_{ickl}u_{id}^{*}u_{ky}u_{ly}\tilde{t}_{dd,yy}\right] + 2\sum_{xijk}\mathrm{Re}\left[t_{ijkc}u_{ix}^{*}u_{jx}u_{kd}^{*}\tilde{t}_{xx,dd}\right] \\
\frac{\partial O(\kappa)}{\partial u_{cd}^{*}} = 2\sum_{xy}\mathrm{Re}\left[\tilde{t}_{xx,yy}\frac{\partial \tilde{t}_{xx,yy}}{\partial u_{cd}^{*}}\right] = 2\sum_{yjkl}\mathrm{Re}\left[t_{cjkl}u_{jd}u_{ky}^{*}u_{ly}\tilde{t}_{dd,yy}\right] + 2\sum_{xijl}\mathrm{Re}\left[t_{ijcl}u_{ix}^{*}u_{jx}u_{ld}\tilde{t}_{xx,dd}\right]
\end{align}
which for all $\{c, d\}$ is obtained in $\mathcal{O}(n^{5})$ once per gradient call.  The $n^{2}$ intermediates are reused for each $\kappa_{ab}$ derivative.  Thus the overall scaling is $\mathcal{O}(n^{5})$.  Using the Wilcox identity~\cite{wilcox1967exponential} the $\{c,d\}$ element of the partial derivative of the unitary $\left(\frac{\partial u}{\kappa_{a,b}}\right)_{cd}$ or its Hermitian conjugate is
\begin{align}
\left(\frac{\partial u}{\partial \kappa_{ab}}\right)_{cd} = \sum_{r}W^{ab}_{cr}u_{rd} \label{eq:partial_u_kab}\\
\left(\frac{\partial u^{*}}{\partial \kappa_{ab}}\right)_{cd} = \sum_{r}-W^{ab}_{rc}u_{rd}^{*}. \label{eq:partial_ustar_kab}
\end{align}
One can simply check that the above derivative is consistent with the $\mathcal{O}(n^7)$ scaling method by multiplying Eq.~\eqref{eq:partial_u_kab} with $\frac{\partial \tilde{t}_{xx,yy}}{\partial u_{cd}}$ and summing over $\{c,d\}$ which returns the appropriate terms in Eq.~\eqref{eq:n7_scaling_deriv}.

Using the infrastructure from the previous section we now introduce a recursive technique for generating the compressed two-body operator:
\begin{enumerate}
\item Starting from the desired operator $T$ we first maximize Eq.~\eqref{eq:opt_max_kappa} to obtain orbital rotation such that the $n_{i}n_{j}$ coefficients are largest in magnitude.  
\item Select out the $n_{j}n_{j}$ coefficients and store them along with the $\kappa$ rotation.  
\item Rotate the operator represented by the diagonal coefficients back to the original basis with the $\kappa$ just obtained from the optimization, and then subtract the tensor from the original generating a remainder.  
\item Repeat steps 1-3 until the norm of the subtraction remainder is below a predefined threshold.
\end{enumerate}
Using this approach the iteration cost of our optimization is never more than the $O(n^{5})$ cost of orbital rotation. Throughout the rest of this paper this recursive fitting procedure will be referred to as the ``unitary compression'' technique.

\section{Results}\label{sec:results}
We examine the performance of the SVD, Takagi, and unitary compression decomposition for the two-body components of unitary coupled-cluster operators.  In this analysis we compare the number of tensor factors versus maximum absolute error and $L_{2}$-norm variation from the true tensor. To be as comprehensive as possible we also document the performance of unitary compression modified to work on hermitian two electron integral tensor in Appendix~\ref{app:eri_decomp}.
We found that for cluster operator compilations the greedy unitary compression technique requires very few tensors to reach sub milliHartree accuracy but suffers a substantial slow down in optimization due to increasing rank of the residual.  All calculations are accomplished with PySCF~\cite{sun2020recent}, OpenFermion~\cite{mcclean2020openfermion}, the Fermionic Quantum Emulator~\cite{rubin2021fermionic}, and a custom implementation of the tensor decomposition schemes.
\subsection{Factorization of Coupled Cluster Doubles}\label{sec:cc_doubles}
\begin{widetext}
\begin{figure*}
    \centering
    \includegraphics[width=8.5cm]{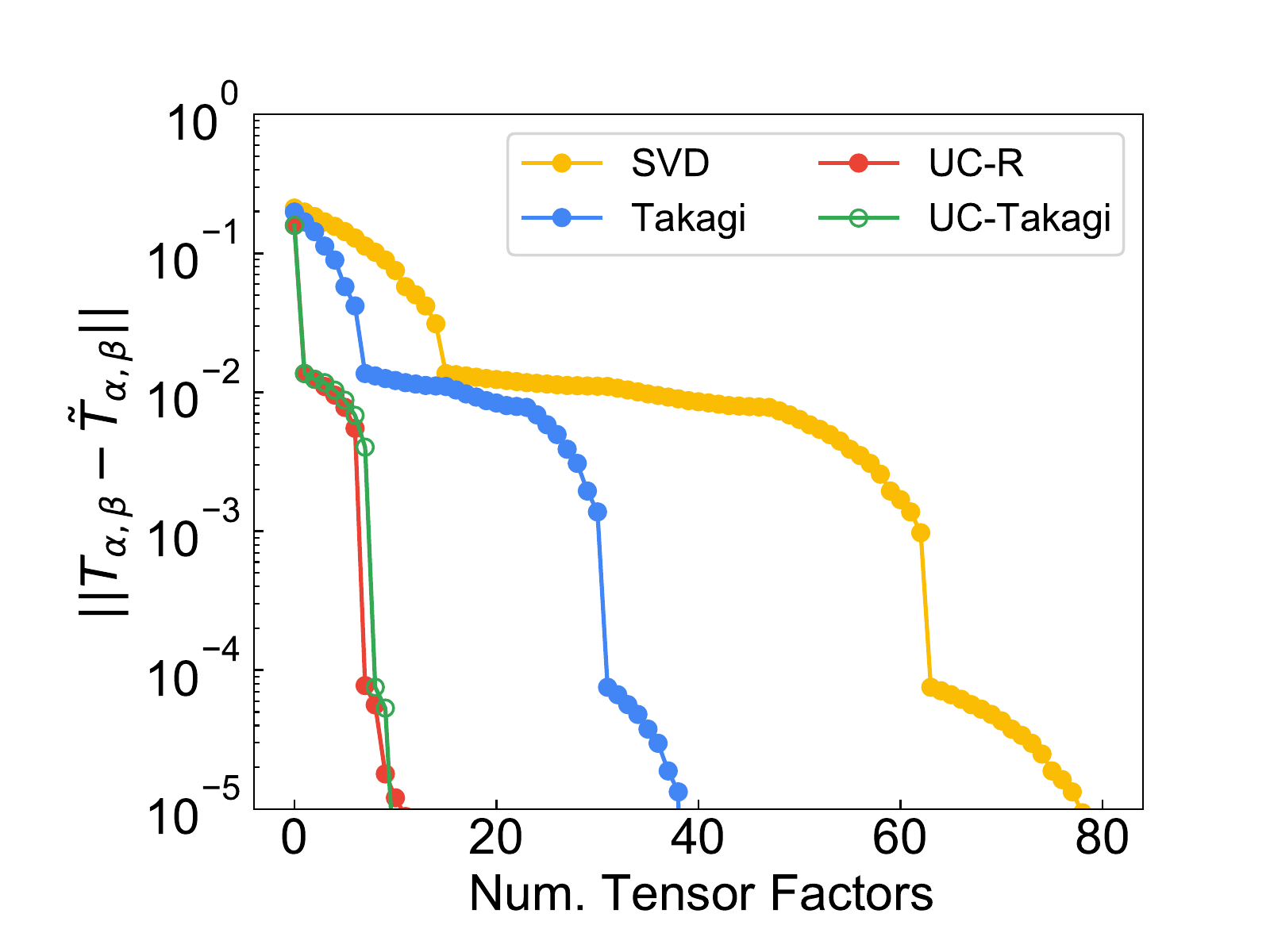}
    \includegraphics[width=8.5cm]{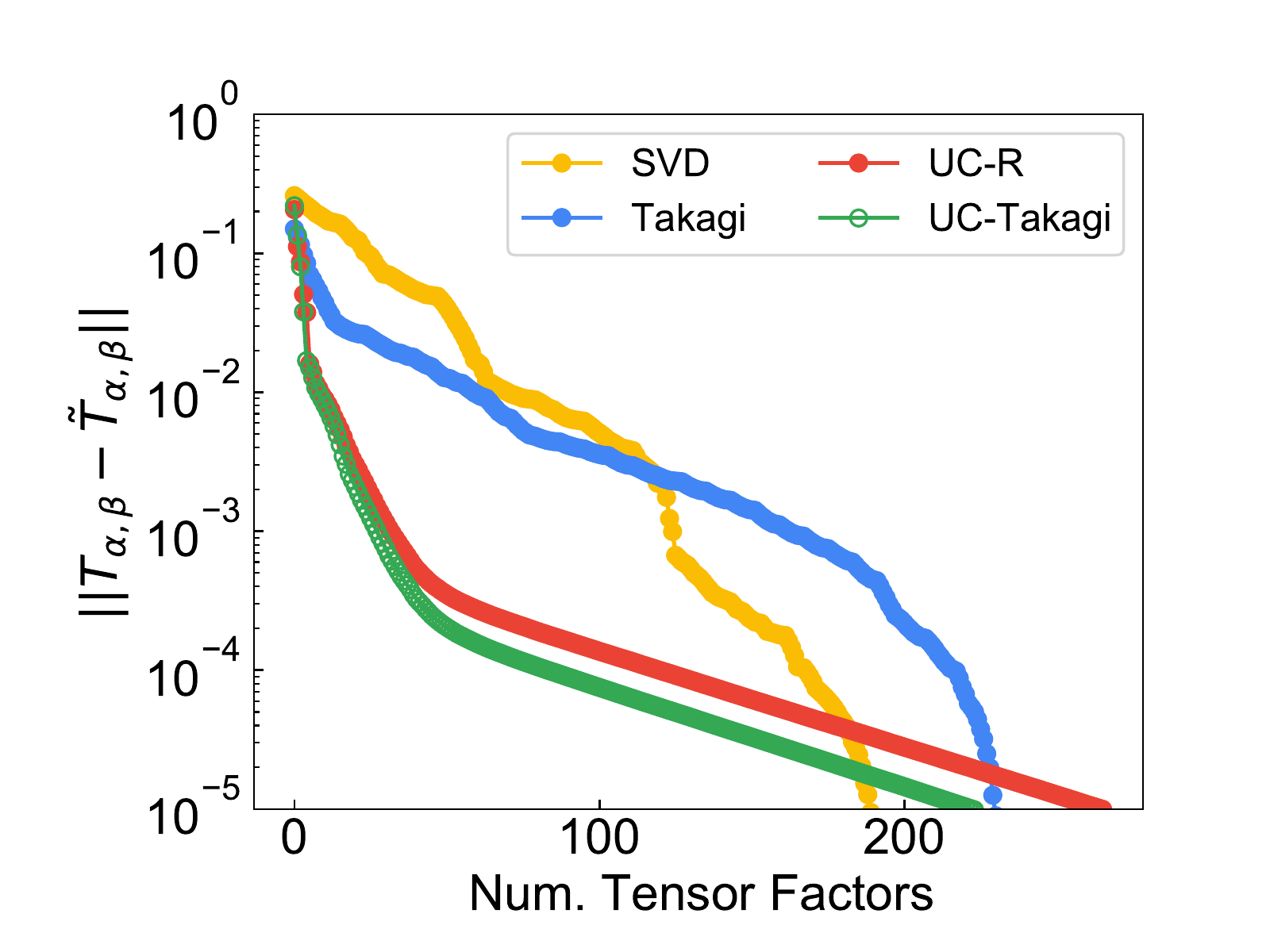}
    \includegraphics[width=8.5cm]{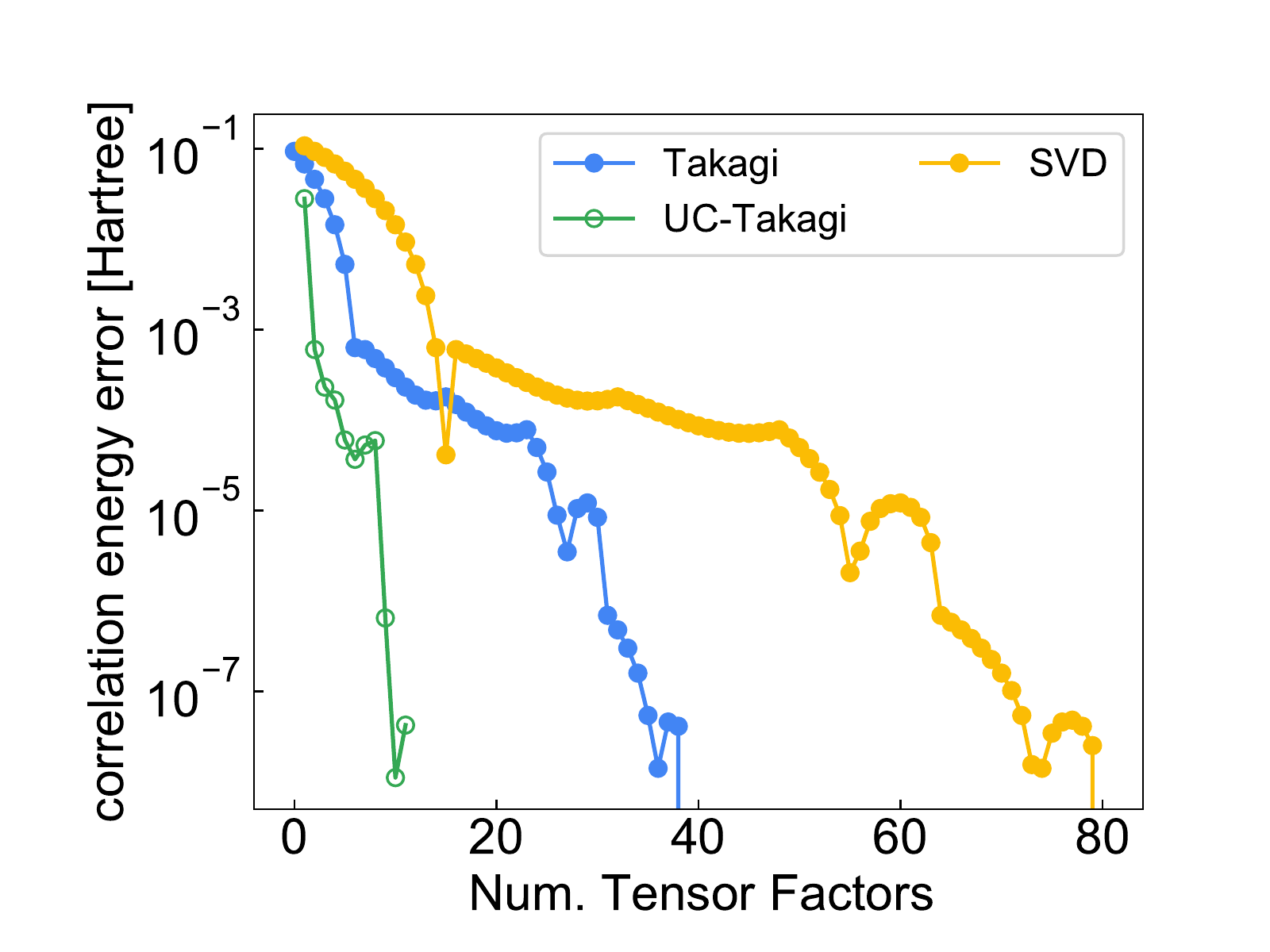}
    \includegraphics[width=8.5cm]{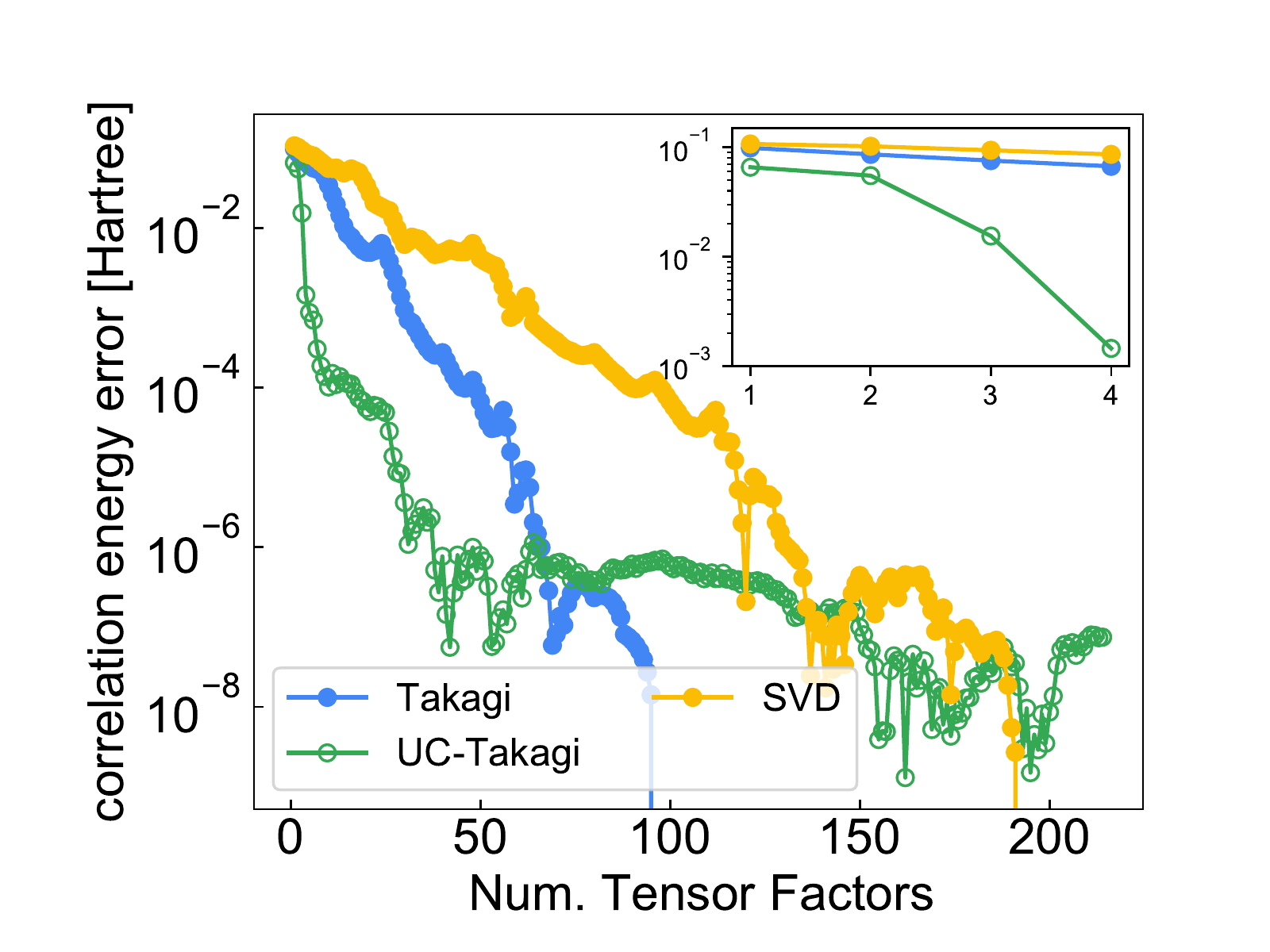}
    \caption{Compression of the $\alpha\beta$ two-body cluster operator for Hydrogen Fluoride (left) in a minimal basis (12 orbitals) and H$_{4}$-linear (right) in a 6-31G basis (16 orbitals).  Unitary compression is labeled with UC.  UC-R in red is the unitary compression with random initial starting states for the orbital optimization.  The UC-Takagi in green is unitary compression starting from from the Takagi decomposition of the remainder matrix.  The SVD decomposition and Takagi decomposition are described in~\ref{app:alternative_decomps}.  Each tensor factor can be implemented with $\mathcal{O}(n)$ depth where $n$ is the number of orbitals. }
    \label{fig:t2_compression}
\end{figure*}
\end{widetext}

The first system we examine is the analytical (via SVD and Takagi) and numerical (unitary compression) decomposition of the two-body generators constructed from classical coupled-cluster singles and doubles (CCSD) solutions.  Given the $T_{2}$ CCSD operator
\begin{align}
T_{2} = \frac{1}{4}\sum_{i,j,a,b}t_{i,j,a,b}a_{a}^{\dagger}a_{i}a_{b}^{\dagger}a_{j}
\end{align}
the unitary generator $\tau_{2}$ can be formed by subtracting the Hermitian conjugate
\begin{align}\label{eq:tau2}
\tau_{2} = T_{2} - T_{2}^{\dagger}
\end{align}
which we decompose into the sum-of-squares form.  We apply the Takagi, SVD, and unitary compression decompositions to two systems, Hydrogen Fluroide in a minimal basis and linear H$_{4}$ in a 6-31G basis both with bond lengths of 1.6 $\mathrm{\AA}$.  In Figure~\ref{fig:t2_compression} we plot the $L_{2}$-norm difference of the $\alpha\beta$-amplitudes tensor and the unitary compressed tensor as a function of the number of tensor factors considered.  For the unitary compression we consider random initialization for the basis rotation coefficients or coefficients obtained from the top vector of the Takagi decomposition. We also plot the convergence of the correlation energy as a function of tensor factors.  For these two systems there is little difference between the unitary compression seeded with the top eigenvector of the Takagi decomposition and random one-body unitaries.  The residual $L_{2}$-norm for the unitary compression technique drops quickly but eventually converges with slower scaling than the initial steps. Despite this slow-down the initial tensors from unitary compression capture enough information such that the correlation energy converges to sub milliHartree levels with respect to CCSD well before the Takagi decomposition and the SVD decomposition.

\begin{figure}[htbp]
    \centering
    \includegraphics[width=8.5cm]{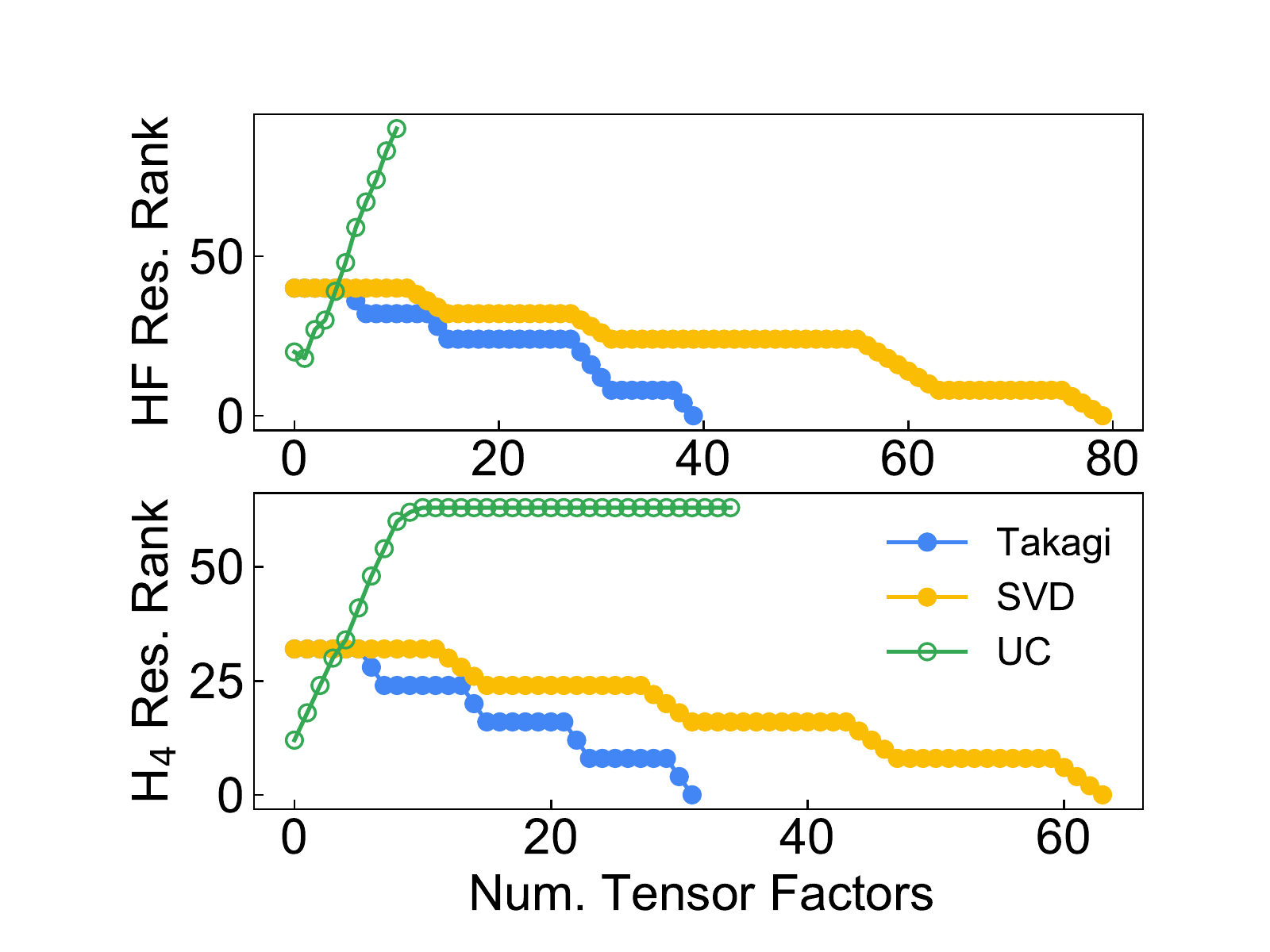}
    \caption{Rank of the residual being fit with Takagi, SVD, and unitary compression (UC).  The Takagi and SVD residual ranks are determined by reconstructing the tensor with a given number of tensor factors ($x$-axis) and computing the rank of the remainder using the Takagi decomposition. The residual rank for unitary compression is the Takagi rank of the residual tensor for the next fitting iteration.}
    \label{fig:infty_norm_rank}
\end{figure}

The cause of the convergence slowdown is likely due to the fact that the nuclear norm of the residual is not being minimized in the greedy procedure.  To show this we plot the rank of the residual being fit by the unitary compression procedure and compare against the residual in the Takagi and SVD case.  The residual in the Takagi and SVD cases are simply the true amplitude tensor minus the reconstructed tensor with a given number of tensor factors.  Using all the tensor factors produces an exact amplitude tensor and thus the residual rank for the Takagi and SVD decomposition asymptotes to zero.  In Figure~\ref{fig:infty_norm_rank} we show the rank of the residuals for Hydrogen Fluoride molecule (HF) and linear H$_{4}$ molecule as a stacked plot.  As expected for Takagi and SVD decompositions the rank of the residual (or remainder tensor) goes to zero.  For the  unitary compression the rank quickly rises to its maximal value.  Once the rank is maximized the convergence of the unitary tensor fitting slows down substantially as seen in previous plots.  Effectively, the unitary compression approach partitions the $\tau_{2}$ coefficient tensor into a low-rank component that captures the majority of the correlation energy and a maximal rank residual component with many small amplitudes.  This suggests to either include the nuclear norm of the residual in unitary compression objective or consider a hybrid scheme where unitary compression is used until the residual rank is maximal and then switch to a Takagi decomposition on the remainder tensor. 

\subsection{Many-body starting states for iterative wavefunction construction}\label{sec:adapt_steps}
To further illustrate the utility of unitary compression we consider starting states for the iterative circuit construction technique ADAPT~\cite{grimsley2018adapt}.  In many classical and quantum algorithms the initial state can vastly change the success probability of the algorithm.  Here we demonstrate a system where ADAPT converges to a state substantially higher in energy than the ground state when initialized with a symmetry preserving Hartree-Fock state and succeeds in finding a low-energy state when an approximation, through unitary compression, to a CCSD initial state is used.  All ADAPT calculations used operator pools of $S_{z}$-adapted two-body operators and numerical optimization was performed with BFGS~\cite{nocedal2006numerical}.  Gradients were obtained through the dynamic programming approach described in Reference~\cite{crooks2019gradients}.  All numerics were performed with the fermionic quantum emulator~\cite{rubin2021fermionic}. 

\begin{figure}[b]
    \centering
    \includegraphics[width=8.5cm]{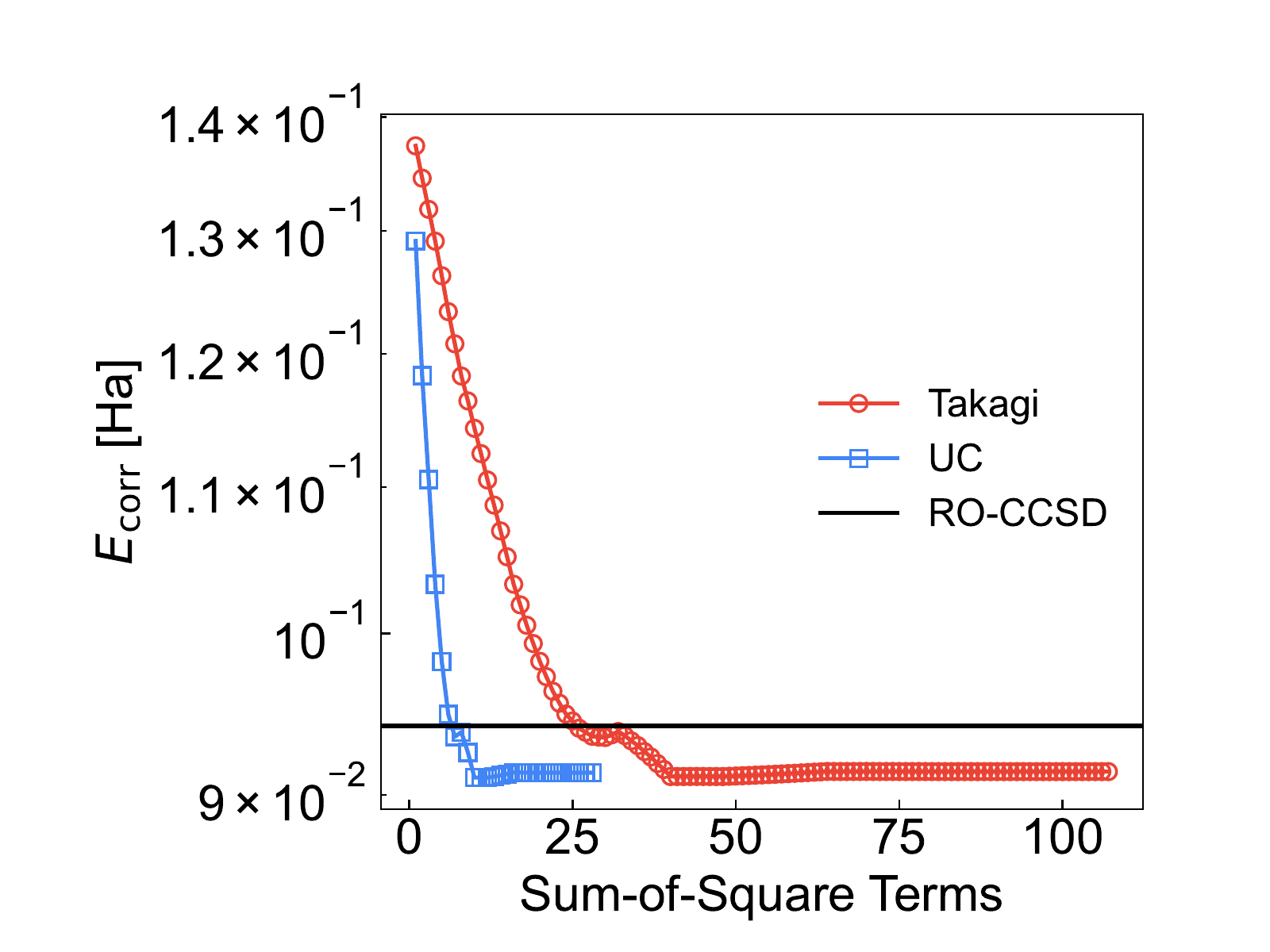}
    \caption{Correlation energy convergence of compiled unitary coupled cluster as a starting state for ADAPT.  Blue is unitary compression and red is the Takagi decomposition. }
    \label{fig:ro_ccsd_corr_conv}
\end{figure}

The system we consider is the triplet ground state of \ce{O2} in a minimal STO-3G basis with a bond distance of 2.55$\mathrm{\AA}$.  Internal stability analysis is performed on all self-consistent-field (SCF) calculations to confirm the SCF solution is not a saddle point.  The restricted-openshell Hartree-Fock (ROHF) wavefunction has less than $1 \times 10^{5}$ overlap with the full configuration interaction (FCI) wavefunction and the unrestricted Hartree-Fock (UHF) wavefunction has approximately 35\% overlap with the FCI wavefunction.  This is because UHF should produce a state that is locally a singlet and triplet on each respective O atom.  Thus the overlap for the full triplet \ce{O2} should be non-trivial. In all cases ADAPT-VQE and CCSD calculations are performed on the full space of ten orbitals and sixteen electrons. 

In Figure~\ref{fig:ro_ccsd_corr_conv} we show that unitary compression can be used to approximate CCSD as a starting state with very few tensor factors.  We compare convergence of the correlation energy captured by unitary compression and Takagi on the $\alpha,\beta$ spin blocks of $\tau_{2}$ defined in Eq.~\ref{eq:tau2} as a function of the number of tensor factors considered.  Similar to the results of the previous section unitary compression substantially outperforms the Takagi decomposition in terms of circuit depth.  To achieve a correlation energy similar to CCSD unitary compression requires six factors whereas Takagi requires twenty five factors.

\begin{figure}
    \centering
    \includegraphics[width=8.5cm]{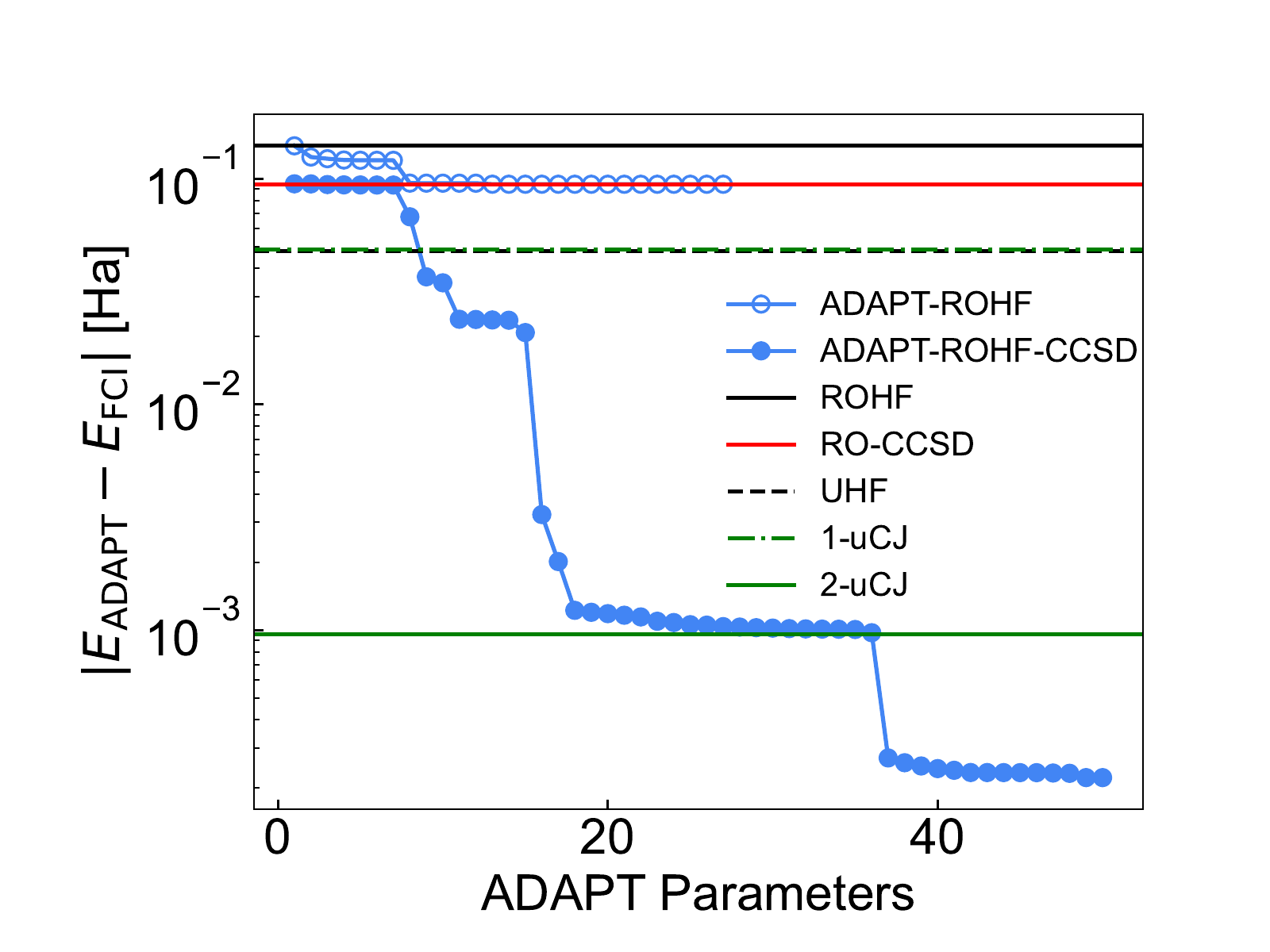}
    \includegraphics[width=8.5cm]{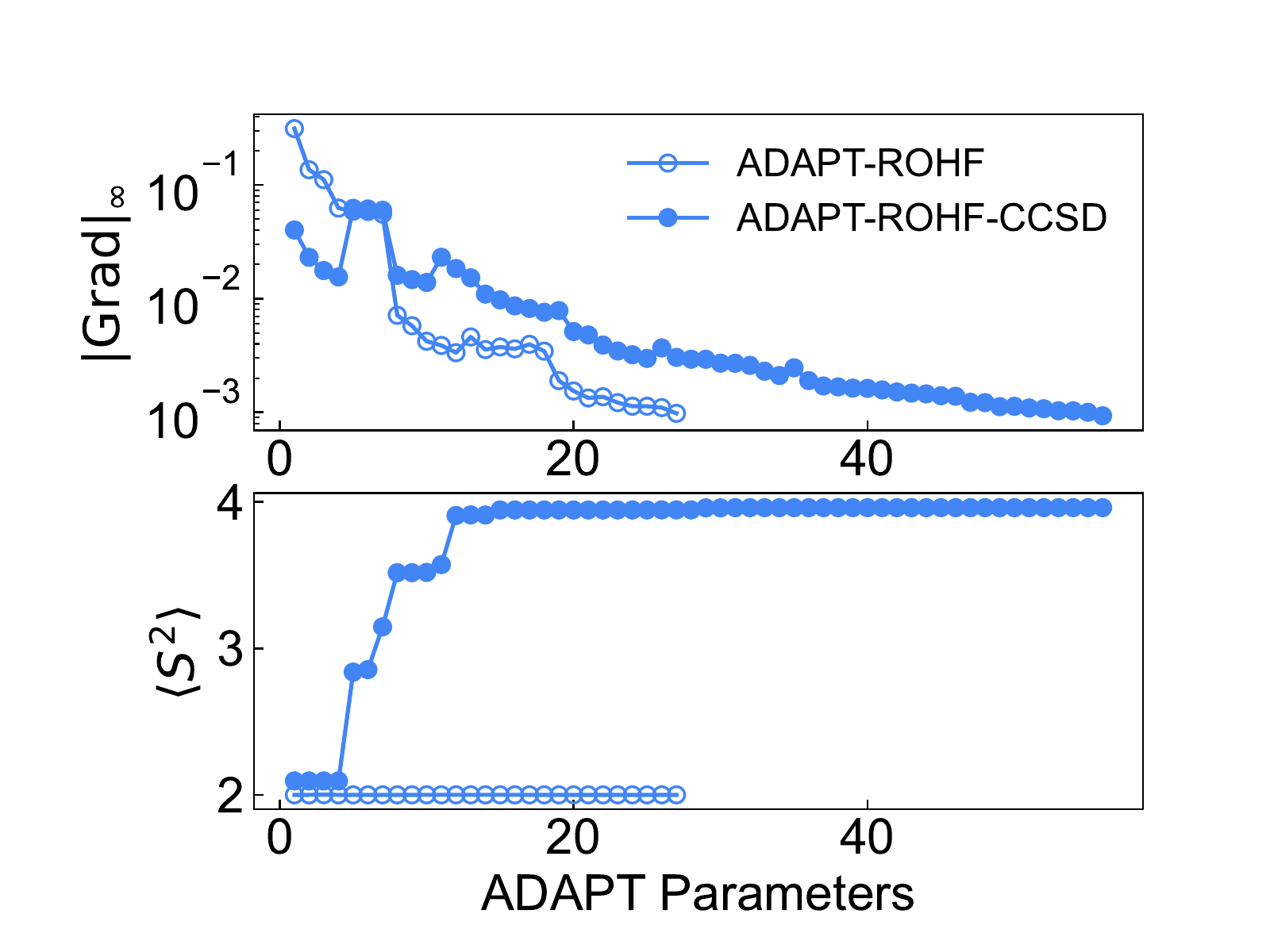}
    \caption{\textit{left}: Convergence of ADAPT for different starting states. Open blue circles correspond to an ROHF starting point and solid blue circles correspond to a RO-CCSD starting point approximated with six unitary compression factors.  Energies for various methods are also shown as reference: ROHF (solid black), UHF (dashed black), RO-CCSD (red), $k=1$ unitary cluster-Jastrow~\cite{matsuzawa2020jastrow} ($k$-uCJ) (dash-dot green), and $k=2$ $k$-uCJ (solid green). \textit{right}: \textit{upper panel:} The maxium absolute deviation of the two-body gradient tensor $g_{pqrs} = \langle\psi \vert \left[a_{p}^{\dagger}a_{q}^{\dagger}a_{s}a_{r}, H\right]\vert \psi \rangle$ after each ADAPT iteration for ROHF starting states and an approximate RO-CCSD starting state.  \textit{lower panel:} $S^{2}$ expectation value after each adapt iteration demonstrating symmetry breaking in ADAPT steps.}
    \label{fig:o2_ccsd_trott_uthc_takag}
\end{figure}

On the left panel of Figure~\ref{fig:o2_ccsd_trott_uthc_takag} we plot the progress of the ADAPT algorithm starting from a Hartree-Fock starting point (ROHF) and a Coupled-Cluster starting point (RO-CCSD) approximated with six unitary compression factors.  In blue open circles we show how ADAPT starting from an ROHF wavefunction fails to converge to a ground state due to the difficulty of finding high quality rotations with a gradient based approach when the system has almost zero overlap the exact ground state. Starting from RO-CCSD approximated with six unitary compression tensors (solid blue circles) ADAPT can succeed, but with substantial circuit depth.  We note here that UHF as a starting point for ADAPT succeeds but with substantial symmetry breaking. The varying performance of ADAPT depending on starting state symmetry breaking hints at the importance of symmetry breaking for the algorithm overall.  For reference we also plot the performance of the $k=1$ and $k=2$ unitary cluster Jastrow ansatz~\cite{matsuzawa2020jastrow} ($k$-uCJ) where the generalized singles term (restricted such that rotations in the $\alpha$ and $\beta$ spin sectors are equivalent) is implemented separately from the pair doubles term in a similar compilation described in Figure~\ref{fig:takagi_circuit_spin_sector}. For O$_{2}$ in a minimal basis $k$-uCJ has $k100$ parameters.

On the right panel of Figure~\ref{fig:o2_ccsd_trott_uthc_takag} we plot the maximum absolute value of the $2$-body gradient set $\{\langle \psi \vert \left[a_{p}^{\dagger}a_{q}^{\dagger}a_{r}a_{s}, H\right]\vert \psi \rangle\}$ as a function of ADAPT iteration.  At each iteration a single spin-adapted two-body operator is added to the wavefunction.  Thus, $S^{2}$ symmetry can be broken by ADAPT. We use $S_{z}$ spin-adapted operators because singlet and triplet two-body operator implementation require Trotterization depending on implementation strategy.  With only $S_{z}$ spin-adapted function there is no way to fix $S^{2}$ expectation values through the ADAPT protocol unless a penalty term is added.  To demonstrate symmetry breaking as a mechanism for ADAPT successfully finding a stationary state we also plot $S^{2}$ as a function of ADAPT iteration.  With the improved starting state of approximate RO-CCSD seventeen ADAPT iterations are required to come within 1 milliHartree of the $k=2$ $k$-uCJ energy. 

\section{Discussion}
We have explored a variety of compilation techniques for implementing many-body fermion dynamics and draw connections between known implementation strategies, such as the SVD or Takagi decomposition, circuit ans\"atze like $k$-uCJ, and a sum-of-squares representation of a generic operator.  All the referenced strategies seek to decompose a many-body operator into a sum-of-squares of normal operators which can be implemented with Trotter error using interleaved Givens rotation networks and Ising interaction networks.  The many-body decomposition schemes based on SVD or Takagi suffer from a rank deficiency in the Ising interaction matrix~\cite{matsuzawa2020jastrow} which can be partially alleviated through full variational relaxation. 

We proposed a strategy for determining a sum-of-squares decomposition of a general two-body operator by numerically searching for a low depth non-orthogonal one-particle bases expansion of the operator.  The greedy numerical decomposition has iteration complexity no worse than a single particle basis transformation.  The decomposition can be applied in energy measurement schemes as in~\cite{yen2020cartan} or for time evolution or ansatz construction. 

The numerical sum-of-squares decomposition clearly outperformed analytical decompositions for approximating a unitary coupled-cluster generator resulting in substantial circuit depth reduction. Thus we can recommend unitary compression as a compilation strategy when the goal is to implement a many-body operator of unitary coupled-cluster form.  Unitary compression can also be applied to many-body interaction terms of higher rank with an appropriate increase in iteration complexity mirroring single particle basis rotation costs.
We demonstrated the use of coupled-cluster compilation via unitary compression as a starting state for ADAPT-VQE.  Without approximate coupled-cluster initial state ADAPT fails to converge to the ground state for triplet \ce{O2}.  Though unitary compression presents a nice starting point it is important to note that $k=1$ $k$-uCJ, which is an instance of a fermionc non-Gaussian state, is efficiently simulable~\cite{shi2018variational, kaicher2021algorithm} and also provides a low depth route for improving the starting state for ADAPT-VQE.

The numerical sum-of-squares decomposition is a generic and useful tool for translating low symmetry many-body operators into quantum circuits and a unifying framework for many of today's algorithms for simulating fermions on near-term and future quantum computers.

\section*{Acknowledgement}
JL thanks David Reichman for support.  We thank William J. Huggins for a detailed reading of the manuscript and discussions.
\bibliography{library}
\onecolumngrid
\newpage

\appendix

\section{Decompositions of the doubles generator}\label{app:alternative_decomps}
\subsection{An indexing prelude}
Given a four tensor there are many ways to perform reshaping into a matrix representation.  The most common reshaping used in quantum chemistry is to use a geminal index for the rows and columns forming a symmetric or antisymmetric matrix.  For example, consider a four index tensor $V_{ijlk}$ associated with a two-electron interaction operator
\begin{align}
V = \sum_{ijkl}V_{(ij),(lk)}a_{i}^{\dagger}a_{j}a_{k}^{\dagger}a_{l}
\end{align}
where we have added parenthesis around the $\{i,j,k,l\}$ indices to signify which pairs are used as geminal indices.  Also note the order reversal for $\{l,k\}$ with respect to the order of the ladder operators.  When we convert the four index tensor to a matrix $M$ by C-order (row-major) reshaping the matrix becomes symmetric and off diagonal elements $\{M_{ij}, M_{ji}\}$ correspond to the coefficients of hermitian conjugate pairs of ladder operator term.  Consider a matrix form of the four-tensor and specifically two off diagonal elements $\{(12,34),(34,12)\}$.  According to our original notation these elements correspond to $V$ operator elements
\begin{align}
V_{(12),(34)}a_{1}^{\dagger}a_{2}a_{4}^{\dagger}a_{3} \;\;,\;\; V_{(34),(12)}a_{3}^{\dagger}a_{4}a_{2}^{\dagger}a_{1} 
\end{align}
which if $V_{(12),(34)} = V_{(34),(12)}^{*}$ are Hermitian conjugates of each other.  Now consider a coefficient tensor without the $\{l, k\}$ label permutation respect to the ladder operator order
\begin{align}
V' = \sum_{ijkl}V'_{(ij),(kl)}a_{i}^{\dagger}a_{j}a_{k}^{\dagger}a_{l}.
\end{align}
A C-order reshaping of the coefficient tensor of $V'$ into a matrix results in a complex symmetric tensor.  The same two off diagonal elements $\{(12,34), (34,12)\}$ correspond to $V'$ operator elements 
\begin{align}
V_{(12),(34)}'a_{1}^{\dagger}a_{2}a_{3}^{\dagger}a_{4} \;\;,\;\; V_{(34),(12)}'a_{3}^{\dagger}a_{4}a_{1}^{\dagger}a_{2} 
\end{align}
which re not Hermitian conjugates of each other. To see how these terms relate to each other we consider the four mode component of the vacuum normal ordered terms
\begin{align}
-V_{(12),(34)}'a_{1}^{\dagger}a_{3}^{\dagger}a_{2}a_{4} \;\;,\;\; -V_{(34),(12)}'a_{1}^{\dagger}a_{3}^{\dagger}a_{2}a_{4} 
\end{align}
and thus for a Hermitian or antihermitian operator $V'$ the coefficients must be equivalent.  Therefore, reshaping without $l,k$ permutation results in a complex symmetric tensor.  Another way to see this is consider a vacuum normal ordered operator $Y$
\begin{align}
Y = \sum_{ijkl}Y_{ijkl}a_{i}^{\dagger}a_{j}^{\dagger}a_{k}a_{l}.
\end{align}
by the anticommutation relations $Y_{ijkl}a_{i}^{\dagger}a_{j}^{\dagger}a_{k}a_{l} = Y_{jilk}a_{j}^{\dagger}a_{i}^{\dagger}a_{l}a_{k}$. Converting both of these operators into the form of $V'$ and considering the two body component
\begin{align}
Y_{ijkl}a_{i}^{\dagger}a_{l}a_{j}^{\dagger}a_{k} \;\;,\;\;Y_{jilk}a_{j}^{\dagger}a_{k}a_{i}^{\dagger}a_{l} \\
Y_{ijkl} = V'_{(il),(jk)} \;\;,\;\;Y_{jilk} = V'_{(jk),(il)}
\end{align}
and thus since $Y_{ijkl}$ is complex, the coefficient tensor of $V'$ reshaped into a matrix using C-ordering is a complex symmetric matrix and thus neither Hermitian or antihermitian.

\subsection{A sum-of-squares decomposition via the SVD}
The decomposition proposed by Motta \textit{et al.} uses a singular value decomposition of the coefficient tensor $A$ of $G$ in order to represent the doubles generator via sum-of-squares.  Expressing $G$ in a charge-charge form
\begin{align}
G =& -\sum_{pq,rs}A^{pq}_{sr}a_{p}^{\dagger}a_{q}^{\dagger}a_{s}a_{r} \nonumber \\
=& \sum_{pq,rs}\tilde{A}^{pq}_{sr}\left(a_{p}^{\dagger}a_{s}a_{q}^{\dagger}a_{r} - \delta_{s}^{q}a_{p}^{\dagger}a_{r}\right)
\end{align}  
suggest to reshape the original $A$ tensor into $\tilde{A}$ such that it is a complex symmetric matrix.  $\tilde{A}$ can then be decomposed by a singular-value decomposition
\begin{align}
\tilde{A} = U \sigma V^{\dagger} \\
\tilde{A}^{ps}_{qr} = \sum_{l}\sigma(l)U_{ps}^{l}V_{qr}^{l}
\end{align}
where index $l$ ranges over all non-zero singular values. $U_{ps}^{l}$ and $V_{qs}^{l}$ are the elements of the $l^{\mathrm{th}}$ column of the left and right singular vectors $U$ and $V$. Defining $\hat{U}_{l}$. and $\hat{V}_{l}$ to be
\begin{align}
\hat{U}_{l} = \sqrt{\sigma(l)} \sum_{ps}U_{ps}^{l}a_{p}^{\dagger}a_{s} \\
\hat{V}_{l} = \sqrt{\sigma(l)} \sum_{qr}V_{qr}^{l}a_{q}^{\dagger}a_{r}
\end{align}
$G$ can be written as 
\begin{align}
G = \frac{1}{4}\sum_{l}\hat{U}_{l}\hat{V}_{l} + \hat{V}_{l}\hat{U}_{l} - \hat{U}_{l}^{\dagger}\hat{V}_{l}^{\dagger} - \hat{V}_{l}^{\dagger}\hat{U}_{l}^{\dagger}.
\end{align}
Each pair of terms can be written as the sum of squares of $\hat{U}_{l} + \hat{V}_{l}$, $\hat{U}_{l} - \hat{V}_{l}$, $\hat{U}_{l}^{\dagger} + \hat{V}_{l}^{\dagger}$, $\hat{U}_{l}^{\dagger} - \hat{V}_{l}^{\dagger}$
\begin{align}\label{eq:svd_z}
G =& \frac{1}{8}\sum_{l}\left[\left(\hat{U}_{l} + \hat{V}_{l}\right)^{2} - \left(\hat{U}_{l} - \hat{V}_{l}\right)^{2} \right. \\
&\left. -\left(\hat{U}_{l}^{\dagger} + \hat{V}_{l}^{\dagger}\right)^{2} + \left(\hat{U}_{l}^{\dagger} - \hat{V}_{l}^{\dagger}\right)^{2} \right].
\end{align}
Just as in~\cite{motta2018low}, we can use the relation for operator $X$
\begin{align}
2 \left(X^{2} - \left(X^{\dagger}\right)^{2}\right) = \left(X - iX^{\dagger}\right)^{2} + \left(X + iX^{\dagger}\right)^{2}
\end{align}
and rewrite $G$ as a sum-of-squares of normal one-body operators~\footnote{For $O \pm iO^{\dagger}$ to be a normal operator for arbitrary $O$ then we must show that it commutes with its adjoint
\begin{align}
\left[O \pm iO^{\dagger}, O^{\dagger} \mp iO\right] =& \left[O, O^{\dagger} \right] + \left[O, \mp iO\right] \nonumber \\
+& \left[ \pm iO^{\dagger}, O^{\dagger}\right] + \left[\pm iO^{\dagger}, \mp iO\right] \nonumber \\
=& \left[O, O^{\dagger}\right] + \mp i\left[O, O\right] \nonumber \\
+&   \mp i\left[O^{\dagger}, O^{\dagger}\right] +  \left[ O^{\dagger}, O\right]   \nonumber
\end{align}
which is zero for any $O$ since an operator always commutes with itself and switching the order in a commutator returns the negative of the original.  Furthermore, the square of each $(O + iO^{\dagger})$ is antihermitian and thus the decomposition of $(O + iO^{\dagger})^{2}$ can be performed and the outer product of eigenvalues will yield pure imaginary numbers.
\begin{align}
((O + iO^{\dagger})(O + iO^{\dagger}))^{\dagger} =& (O^{2} - (O^{\dagger})^{2} + iOO^{\dagger} + i O^{\dagger}O)^{\dagger} \nonumber \\
=& -O^{2} + (O^{\dagger})^{2} - iOO^{\dagger} - i O^{\dagger}O \nonumber \\
= -((O + iO^{\dagger})(O + iO^{\dagger})). \nonumber
\end{align}}
\begin{align}
S_{l} =& U_{l} + V_{l} \;\;,\;\;D_{l} = U_{l} - V_{l} \\
\tilde{G} =& \frac{1}{16}\sum_{l}\left[\left(S_{l} + i S_{l}^{\dagger} \right)^{2} +\left(S_{l} - iS_{l}^{\dagger} \right)^{2} \right. \\
&\left. + \left(D_{l} + i D_{l}^{\dagger}\right)^{2} + \left(D_{l} - i D_{l}^{\dagger}\right)^{2}\right].
\end{align}
At no point did we make use of a restriction on the original form of $A$-tensor coefficients other than their antisymmetry in upper and lower indices which is a point of differentiation from the proof  in~\cite{motta2018low}.

\subsection{A sum-of-squares decomposition via the Takagi decomposition}
Another decomposition that is similar to the SVD is the one used in the context of motivating the unitary Jastrow coupled-cluster ansatz~\cite{matsuzawa2020jastrow}.  In this decomposition the coefficient tensor $A$ of $G$ is formed into a super-matrix (just as in the SVD approach) where each row column is indexed by the composite indices $ps$ and $qr$. This matrix is a complex symmetric matrix which can be decomposed via the Takagi~\cite{PhysRevA.94.062109} decomposition
\begin{align}
\tilde{A} =& U \sigma U^{T}\\
\tilde{A}_{ps,qr} =& \sum_{l}\left(\sqrt{\sigma(l)}U^{l}_{ps}\right)\left(\sqrt{\sigma(l)}U^{l}_{qr}\right) \\
=& \sum_{l}V^{l}V^{l}
\end{align}
where $U^{T}$ is the transpose of $U$ and $\sigma$ is a diagonal matrix. Defining the operator
\begin{align}
\hat{V}_{l} = \sum_{l}\left(\sqrt{\sigma(l)}U^{l}_{ps}\right)a_{p}^{\dagger}a_{s}
\end{align}
the generator can be expressed as
\begin{align}
G = \sum_{l}\hat{V}_{l}\hat{V}_{l}.
\end{align}
$\hat{V}_{l}$ is not a normal operator but can be represented by a linear combination of normal operators
\begin{align}\label{eq:takagi_z}
\hat{V}_{l}^{\pm} = \left(\hat{V}_{l} \pm i V_{l}^{\dagger}\right) \;\;,\;\;G= \frac{1}{4}\sum_{l}\left( \left(\hat{V}_{l}^{+}\right)^{2} + \left(\hat{V}_{l}^{-}\right)^{2}\right)
\end{align}
which are normal one-body operators.
\section{Electron Repulsion Interaction Decomposition}\label{app:eri_decomp}
\begin{figure}[h]
    \centering
    \includegraphics[width=8.5cm]{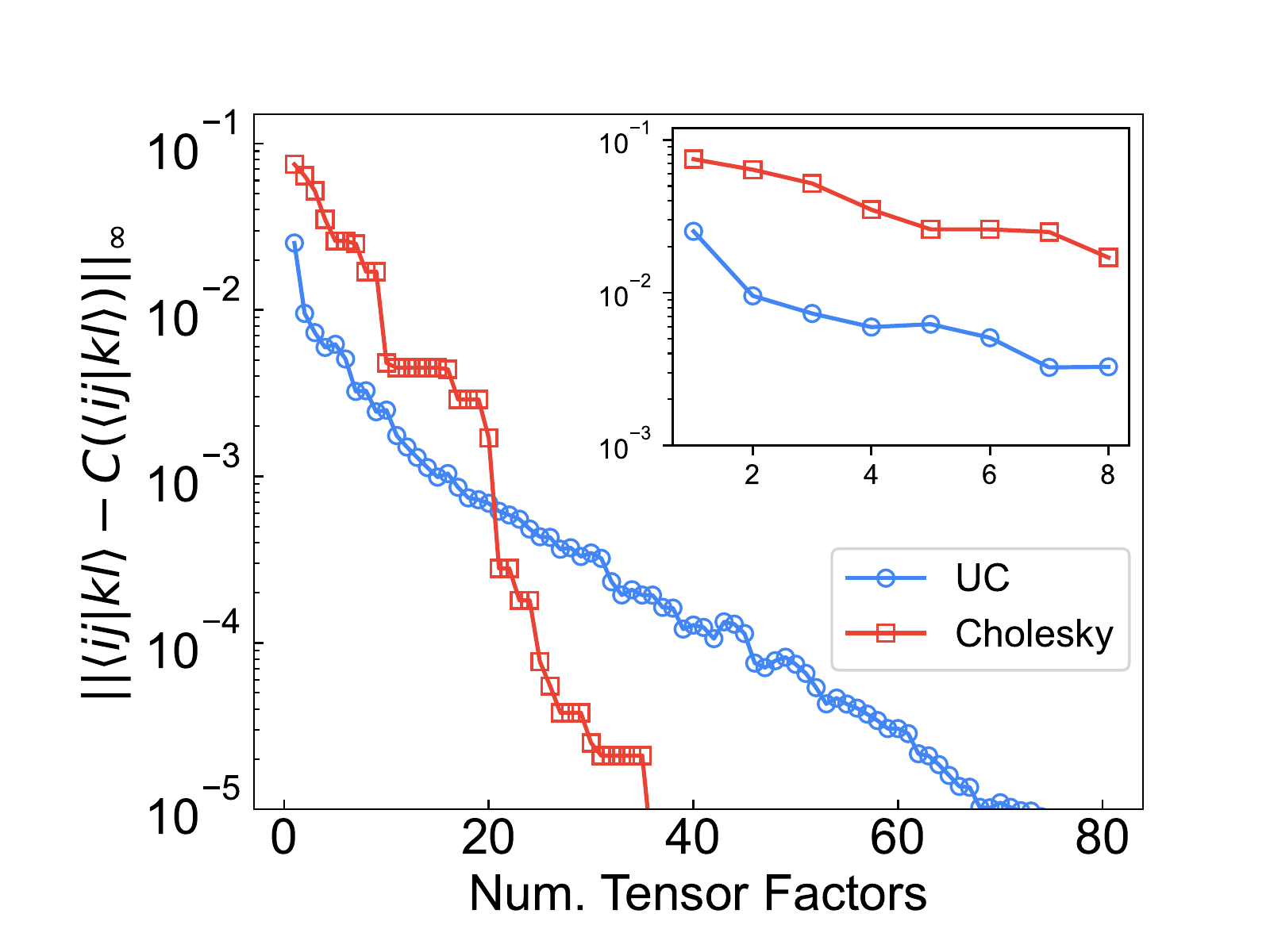}
    \includegraphics[width=8.5cm]{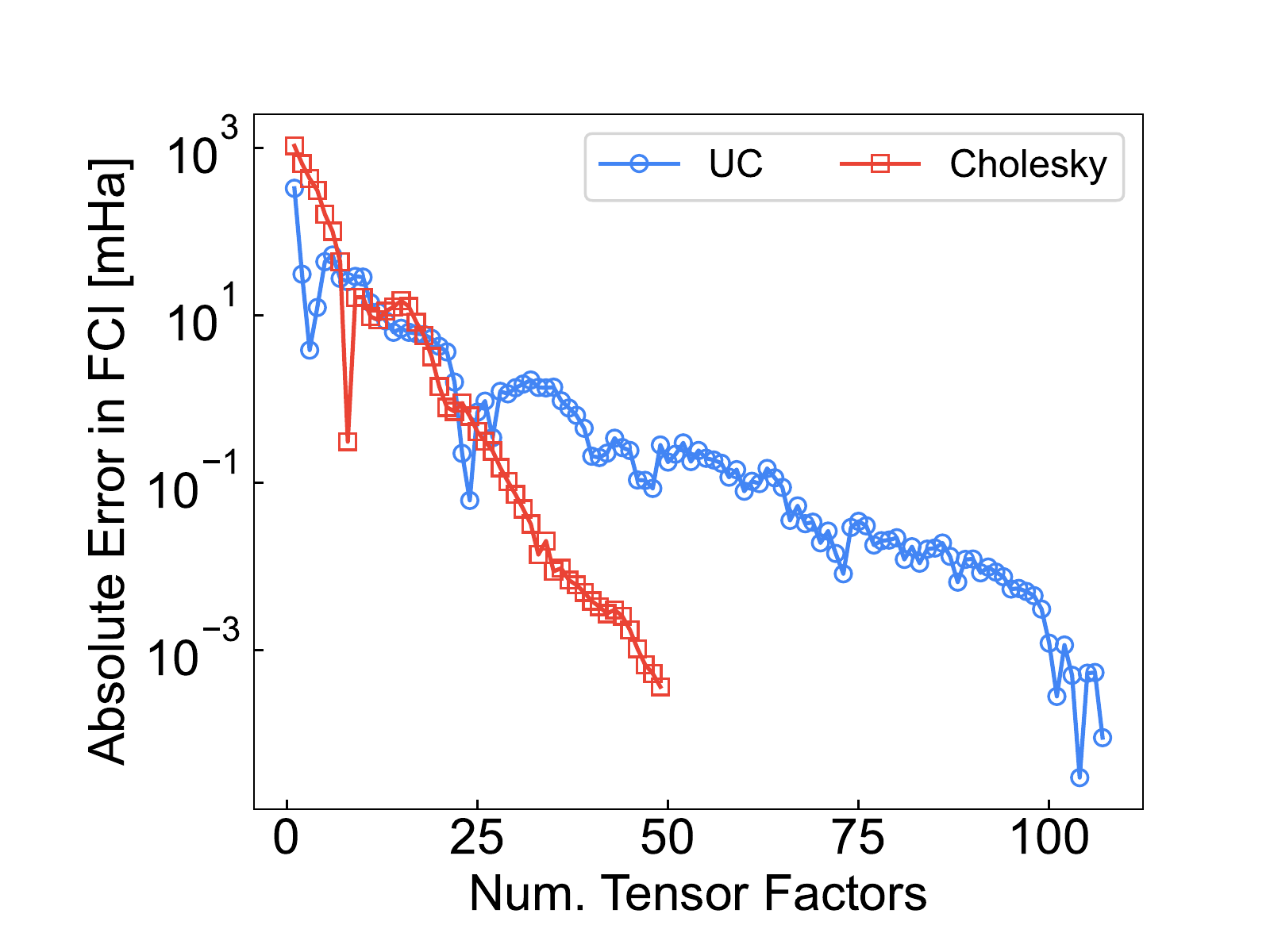}
    \caption{\textit{left} Maximum absolute deviation (MAD) of the unitary compressed two-electron integral tensor of Naphthalene (10 orbital $\pi$ system) obtained from canonical Hartree-Fock orbitals in a cc-pVDZ basis compared against the Cholesky factorized two-electron integral tensor. Inset plot is the MAD for the first eight tensor factors. \textit{right} Absolute difference from FCI energy computed with untruncated two-electron integrals. In both \textit{left} and \textit{right} plots we restrict the compression to real unitary rotations.}
    \label{fig:napthal_mad}
\end{figure}

We consider unitary compression on the two-electron integral tensor for the $\pi$-system of Naphthalene computed in a cc-pVDZ basis at the geometry from Reference~\cite{mullinax2018analytic}. We demonstrate that unitary compression achieves similar performance to the least-squares fitting of Reference~\cite{cohn2021quantum} but does not suffer from computational slow-downs with an increase in the number of fitting tensors.  We modify the unitary compression scheme to fit hermitian operators and optimize over the space of spatial orbital rotations. In Figure~\ref{fig:napthal_mad} we plot the maximum absolute deviation of the unitary compressed two-electron integral tensor.  The unitary compression protocol is performed until the residual $L_{2}$-norm is below $1.0\times10^{-5}$.  This bound is selected based on the convergence of second order M{\o}ller-Plesset perturbation theory (MP2) and coupled-cluster with singles and doubles (CCSD) with truncated low-rank factorization of the two-electron integrals for the large metal-organic catalyst FeMoCo~\cite{lee2020even}.  The performance of unitary compression is compared against a Cholesky factorization of the two-electron integral matrix. The Cholesky decomposition is performed via an SVD. Unitary compression succeeds in lowering the maximum absolute deviation over the Cholesky decomposition but only up until twenty tensor factors.  On the right hand side of Figure~\ref{fig:napthal_mad} we plot the absolute error in exact diagonalization (FCI) energies computed with truncated two-electron integral operators.  The vastly different energies from FCI indicate that though unitary compression is lowering the maximum absolute deviation of the two-electron integral tensor the resulting Hamiltonian is very different. This brings into question whether unitary compression is a useful technique for compressing operators derived from the Coulomb kernel. 

\end{document}